# Understanding Intention to Adopt Smart Thermostats: The Role of Individual Predictors and Social Beliefs across Five EU Countries

Mona Bielig[1,2][a], Florian Kutzner[1][b], Sonja Klingert[3][c] and Celina Kacperski[1,2][d]
[1]*Seeburg Castle University, Seeburgstraße 8, 5201 Seekirchen am Wallersee, Austria*
[2]*University of Konstanz, Universitätsstraße 10, 78464 Konstanz, Germany*
[3]*University of Stuttgart, Keplerstraße 7, 70174 Stuttgart, Germany*
*mona.bielig@uni-seeburg.at, florian.kutzner@uni-seeburg.at, sonja.klingert@ipvs.uni-stuttgart.de, celina.kacperski@uni-seeburg.at*

Abstract: Heating of buildings represents a significant share of the energy consumption in Europe. Smart thermostats that capitalize on the data-driven analysis of heating patterns in order to optimize heat supply are a very promising part of building energy management technology. However, factors driving their acceptance by buildings' inhabitants are poorly understood although being a prerequisite for fully tapping on their potential. In order to understand the driving forces of technology adoption in this use case, a large survey (N = 2250) was conducted in five EU countries (Austria, Belgium, Estonia, Germany, Greece). For the data analysis structural equation modelling based on the Unified Theory of Acceptance and Use of Technology (UTAUT) was employed, which was extended by adding social beliefs, including descriptive social norms, collective efficacy, social identity and trust. As a result, performance expectancy, price value, and effort expectancy proved to be the most important predictors overall, with variations across countries. In sum, the adoption of smart thermostats appears more strongly associated with individual beliefs about their functioning, potentially reducing their adoption. At the end of the paper, implications for policy making and marketing of smart heating technologies are discussed.

## 1. INTRODUCTION

Around 40% of energy in the EU is consumed in buildings, and of building energy consumption about 80% is used for heating[1]. In order to meet the 2030 target of a 55% reduction in emissions compared to 1990, which heavily involves the building sector (European Environment Agency (EEA), 2021), there has been a push for the adoption of smart home technologies, including smart energy management technologies (European Commission, 2022). A smart thermostat, a specific type of smart heating technology (SHT), connects to the existing heating system and detects behavioral patterns of residents, in some cases allows for smart controls, and can save up to 30% energy, depending on the type (Lu et al., 2010; Wang et al., 2020).

---

[a] https://orcid.org/0000-0001-7535-8961
[b] https://orcid.org/0000-0000-0000-0000
[c] https://orcid.org/0000-0003-0653-003X
[d] https://orcid.org/0000-0002-8844-5164

[1] https://energy.ec.europa.eu/topics/energy-efficiency/energy-efficient-buildings/energy-performance-buildings-directive_en

In principal, smart thermostats work in a loop of detection of behavioral patterns through sensors, and potentially with the prediction of external events or temperature, in order to predict dynamic heating needs under comfort constraints and then provide optimal requirement to heating supply, as shown in Figure 1 (Haji Hosseinloo et al., 2020).

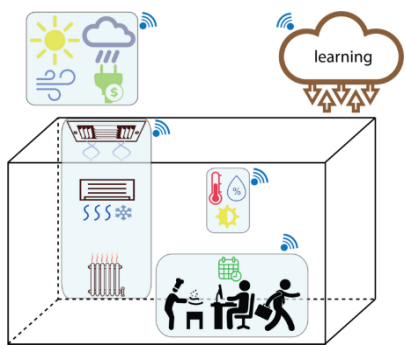

Figure 1: Design of smart thermostats (Haji Hosseinloo et al., 2020)

Smart home technology may not yet be widely perceived as mainstream (Chang & Nam, 2021), and its rate of adoption has been characterized as relatively slow (Marikyan et al., 2019). However, market trends suggest a gradual expansion in the sector (Sovacool & Furszyfer Del Rio, 2020). Eurostat figures from 2022 reveal that about 10% of European households have incorporated smart home technologies, including but not limited to energy management systems.

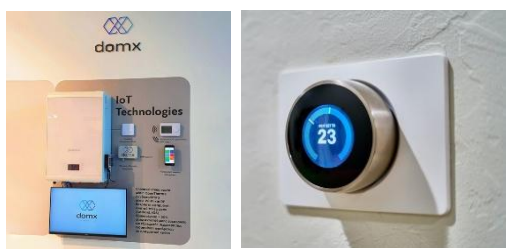

Figure 2: Left: Smart heating technology used as the basis for technical design. Right: General technology picture shown to participants.

Notably, the widespread rollout of smart thermostats not only requires the availability of technical equipment, but also households' willingness to adopt the smart thermostats, and a consent to give up some control. In order to test technology acceptance, a real life example was used (see Figure 2) to describe the way that a smart thermostats works to non-scientific participants of a survey. In the following, we will first describe the underlying theoretical models and related work on predictors for smart thermostat adoption, resulting in our proposed model and hypotheses. We will then describe our study design to understand intentions to adopt smart thermostats in five European Countries, followed by an overview on our most important results. We end with discussions and conclusions of our work, including limitations and strengths as well as policy implications.

## 2. RELATED WORK

Previous acceptance research has focused mainly on smart home technologies in general (for a review, see Li et al., 2021), with research on smart heating technologies in particular under-represented. This is problematic, as behavioral flexibility in heating is lower than for other appliances (Spence et al., 2015). The existing research on acceptance of smart heating technology concentrates on individual and technical factors (e.g., Girod et al., 2017); social drivers for acceptance are either absent or results inconsistent (Große-Kreul, 2022). As both the diffusion of innovative technologies, and pro-environmental decisions have been shown to be driven by social aspects (Fritsche et al., 2018; Rogers, 2003), the current research deepens the understanding of acceptance of a smart heating technology, smart thermostats, combining individual and social aspects in one model.

### 2.1. UTAUT and smart technology acceptance

The Unified Theory of Acceptance and Use of Technology (UTAUT) (Venkatesh et al., 2003) was developed to summarize eight different technology acceptance models and has been successfully employed across multiple contexts such as mobility, IoT in health care or mobile payment (Abrahão et al., 2016; Arfi et al., 2021; Nordhoff et al., 2021). It was extended to the UTAUT2 (Venkatesh et al., 2012) to better align with a consumer context. The full model includes seven predictors for behavioral intention to use a technology, which only then translates into actual user behavior. The seven predictors are *performance expectancy, effort expectancy, social influence, facilitating conditions, hedonic motivation, price value* and *habit*.

While the UTAUT[2], or sometimes a subset of its predictors, have been studied in the context of acceptance of different smart home technologies, the predictive capacity of its components has been inconsistent across studies, with the exception of performance expectancy, which has repeatedly been shown to best predict intention to adopt, both for smart energy technologies in general (Gimpel et al., 2020), and for smart thermostats in particular (Ahn et al., 2016; Girod et al., 2017; Große-Kreul, 2022; Mamonov & Koufaris, 2020).

Effort expectancy, that is the perceived ease of use, was amongst the strongest predictors for behavioral intention to adopt smart energy technologies, i.e. energy-saving technologies comprising sensors and automatic control in a Danish sample (Billanes & Enevoldsen, 2022), and smart homes in a sample from Jordan (Shuhaiber & Mashal, 2019). However, it showed no effects for the intention to adopt smart thermostats in an US sample (Ahn et al., 2016). Hedonic motivations were important for the adoption of smart thermostats in a German sample (Girod et al., 2017), while in a different German study, they were irrelevant (Große-Kreul, 2022). Similar inconsistencies can be found for price value, which had no effect on intention to adopt a smart thermostat (Girod et al., 2017), but was found relevant in a discrete choice experiment (Tu et al., 2021).

The UTAUT also includes the factor social influence, i.e., the belief that important others think an individual should use the technology, which will be discussed in the next section (2.2). Beyond the UTAUT, other technology acceptance factors such as privacy concerns or compatibility have been examined in studies on intention to use or adopt smart thermostats, smart homes, or smart meters.

## 2.2. Social beliefs as predictors of smart technology acceptance

Within the UTAUT, the factor of "social influence" depicts the belief that important others think an individual should use the technology (Venkatesh et al., 2003). Social influence is thus understood as a social norm in the sense of the Theory of Planned Behavior (TPB; Ajzen, 1985). This type of norm describes an individual's perception of what others expect them do and reflects the normative belief or social pressure of 'ought' of (important) others, which is also labeled as an injunctive norm (Cialdini, 2007; Cialdini et al., 1990; Göckeritz et al., 2009; Rivis & Sheeran, 2003). Studies for acceptance of smart thermostats that included social influence found varying effects (Ahn et al., 2016; Billanes & Enevoldsen, 2022; Gimpel et al., 2020; Girod et al., 2017): While it has predicted smart thermostat adoption intention in Germany (Große-Kreul, 2022) and smart meter adoption in Brazil (Gumz et al., 2022), there were very small or no significant effects in other studies (Ahn et al., 2016; Gimpel et al., 2020; Girod et al., 2017). In the two studies in which social influence had the strongest effect on behavioral intention to adopt smart thermostats and smart meters (Gumz et al., 2022), the operationalization included additional aspects less reminiscent of an injunctive norm, such as sheer perceptions of presence of smart thermostats in media, or recommendations by the government. An overview is given in the supplementary material of our article (SM1)[3].

The perception of what other do or believe corresponds to the psychological social norm approach (Berkowitz, 2004), which differentiates between injunctive and descriptive norms. While injunctive norms capture the priorly mentioned normative belief of social pressure, descriptive norms refer to an individual's belief about the prevalence of a behavior, i.e. of what is "normal", therefore the perception of others' own attitudes and behaviors in the domain (Cialdini, 2007; Rivis & Sheeran, 2003). Further, research shows that descriptive and injunctive norms might differ in their effects in changing behavior (e.g. Park & Smith, 2007; White et al., 2009). As prior research failed to demonstrate an effect of *injunctive social norms* for smart thermostat uptake, e.g. in Ahn et al. (2016) or Girod et al. (2017), we aim to investigate *descriptive social norm perceptions* as a possible driver of technology acceptance for smart heating.

Beyond descriptive social norms, there are several other social beliefs that have been shown to influence pro-environmental behavior, which might complement or interact with social norms to promote the adoption of smart thermostats. Prominently, the social identity model of pro-environmental action (SIMPEA) (Fritsche et al., 2018) depicts the influence of *social identity* processes for pro-environmental behavior, also related to the adoption of "green" technologies. Relevant predictors are *collective efficacy* beliefs, i.e. people's beliefs in the effectiveness of their combined ability to achieve goals, and social identification, i.e. the degree to which relevant group memberships are considered important for the individual. Further, generalized *social trust*, referring to general trust in others across groups, and trust in the state, i.e. government and institutions were found decisive for intentions of pro-environmental behavior (Cologna & Siegrist, 2020), and individual energy-saving behavior (Caferra et al., 2021). We therefore explore whether these social beliefs might drive the adoption of smart heating technology. The following section will explicate the definition of our predictors, the prior findings we build them on, then form our hypotheses.

---

[2] We will further refer to the UTAUT-model comprising research based on both UTAUT and UTAUT2.

[3] Supplementary material can be found under https://osf.io/ba2vf/?view_only=986065e170584cad9098d0a2937e216b

## 3. MODEL PREDICTORS AND HYPOTHESES

We examine predictors of the intention to adopt smart heating technology, specifically combining individual beliefs incorporated in the UTAUT, and different social beliefs which have been shown to affect pro-environmental behavior decisions, energy saving intentions and uptake of smart technologies in prior research. Our goal is to examine whether social beliefs can explain the intention to adopt smart heating technology *in addition to* beliefs about technical aspects. We decided to exclude habit, facilitating conditions and hedonic motivation from our model, as no experience with the technology is expected, no active usage is required, and no additional infrastructure beyond the thermostat itself is necessary.

*Behavioral intention (BI)* is our key dependent variable, reflecting the willingness to adopt smart heating technologies, and is the strongest predictor of technology use, especially for technologies with limited consumer experience (Venkatesh et al., 2012). To explain behavioral intention, we consider the following predictors, both based on literature of technology adoption and wider pro-environmental behaviors:

*Performance expectancy (PE)* refers to the perceived usefulness of the technology in achieving specific goals and was found to be a significant predictor of intention to adopt smart energy technologies in multiple studies (Venkatesh et al., 2012; Gimpel et al., 2020; Ahn et al., 2016). *Effort expectancy (EE),* another core UTAUT construct, captures the perceived ease of use and technical efficacy of the technology (Venkatesh et al., 2003), and evidence suggests it plays a role in smart energy technology adoption (Billanes & Enevoldsen, 2022; Ahn et al., 2016). *Price Value (PV)* is the perceived trade-off between the cost of technology and its benefits, which has been shown to influence technology adoption (Venkatesh et al., 2012; Tu et al., 2021).

*Social norms (SN)* reflect the influence of others' perceived behavior on individual intentions. Descriptive neighborhood social norms have been found to influence pro-environmental behaviors (Farrow et al., 2017; Allcott, 2011). *Social identification (SI)*, can moderate this effect of social norms on pro-environmental behavior, i.e. it interacts with the originally expected influence of social norms and might modify it (Cialdini & Jacobson, 2021; Masson & Fritsche, 2014). *Collective efficacy (CE)*, or the belief in the collective ability to achieve environmental outcomes, is another key predictor of pro-environmental behavior (Bandura, 2000; Wang, 2018), and can compensate for low individual efficacy (Jugert et al., 2016).

Lastly, *trust* is a significant predictor of pro-environmental behavior, encompassing both general *trust in people (TP)* and *trust in state institutions (TS)*. Higher trust in others and in institutions has been shown to influence pro-environmental behaviors, particularly energy-saving intentions (Cologna & Siegrist, 2020; Caferra et al., 2021). Demographic variables like age and gender are also considered as control variables, given their influence on smart home adoption (Shin et al., 2018; Sovacool et al., 2020).

This leads us to the following Hypotheses:

H1  *Performance expectancy has a positive effect on BI to adopt smart heating technology.*

H2  *Effort expectancy has a positive effect on BI to adopt smart heating technology.*

H3  *Price value has a positive effect on BI to adopt smart heating technology.*

H4  *Social norms within the neighborhood have a positive effect on BI to adopt smart heating technology.*

H5  *Social identification moderates the effect of social norms, with stronger effects on BI in case of higher neighborhood identification.*

H6  *Collective efficacy beliefs have a positive effect on BI to adopt smart heating technology.*

H7  *General trust in people has a positive effect on BI to adopt smart heating technology.*

H8  *General trust in state has a positive effect on BI to adopt smart heating technology.*

## 3.1. Model

Figure 3 depicts our research model, including both the structural and measurement model. Predictors are split into 'individual (left) and 'social' (right) beliefs.

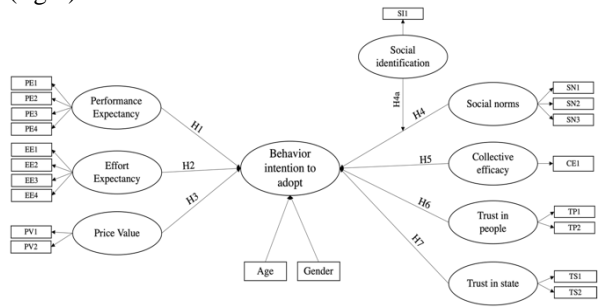

Figure 3: Research model, including both structural and measurement model. All item abbreviations correspond to the item codes in the supplementary material.

## 4. METHOD

Our survey consisted of items related to a smart heating thermostat, its acceptance and the individual and social beliefs that might influence intention to adopt. The survey started with an introduction, a short description of a smart heating device together with an image (see Figure 1), and consent procedures. Afterwards, participants were asked to answer all items included in the questionnaire, ending with demographic details[4].

## 4.1. Sample description

Our sample of N = 3227 was recruited by a professional panel provider and stratified by age and gender for five European countries (Austria, Belgium, Estonia, Germany, Greece). Renumeration was based on the provider's usual rates. Data was gathered through an online survey link between 21.07.2021 and 10.08.2021. Data collection was anonymous and in line with the ethical guidelines of the DGPs (DPGs, 2016). All items were translated into each country's native language by native speakers and back-translated to check for accuracy; each translation was reviewed by a researcher with a native speaker multiple times.

We excluded participants who did not finish the survey, failed the attention check, had an average relative speed index of >2, and used the careless package (Yentes & Chevallier, 2021) to exclude participants with a longstring > 25[5]. This led to a final sample of N = 2250 (51.3% women) from five different countries, with Austria N = 465 (49.7% women), Belgium N = 414 (52.9% women), Estonia N = 510 (52.5% women), Germany N = 425 (50.6% women), and Greece N = 436 (50.7% women). Overall sample size, distributed similarly between countries, was chosen to enable both an overall structural equation modeling (SEM, Hair et al., 2021), as well as country-specific analyses[6]. Participants' distribution across age brackets was 21.7% 18-30 years old, 34.6% between 30-50 years old, 42.1% between 50-70 years old, and 0.4% over 70 years old. Regarding their living situation, 1393 participants (62%) owned their home, while 853 participants (38%) indicated to rent their living space. Most survey participants lived in households between two to four people (87%), with 37% of the households having children. 34% of participants reported heating with a boiler, 22% reported district heating, 17% reported electric heating, and 27% reported using other ways of heating.

## 4.2. Measurements and scales

We examined predictors of acceptance of smart heating technology, specifically combining in one model UTAUT predictors and social beliefs that have been shown in the past to affect technology adoption and/or pro-environmental decisions, in-line with the model shown in Figure 1. Items were surveyed on a seven-point Likert scale ranging from "strongly disagree" to "strongly agree" unless indicated otherwise. We measured Behavioral Intention with four items, e.g., "If I had the opportunity, I would opt for a smart heating appliance" (Abrahão et al., 2016; Venkatesh et al., 2003). Performance Expectancy was assessed with four items, e.g.,"I believe by using such a smart appliance, I would save a meaningful amount of greenhouse gases" (Girod et al., 2017; Venkatesh et al., 2003). Effort Expectancy was measured using four items, e.g., "I believe that using such smart heating control would be: difficult – easy" on a 7-point scale (Venkatesh et al., 2012, 2003). Price Value included two items, e.g., "Such a smart heating appliance is good value for the money" (Venkatesh et al., 2012). Collective Efficacy was measured with one item, "If a large portion of the population used the smart device, we would have a positive effect on society and the climate" (Chen, 2015; Wang, 2018). Social Norms were assessed with three items, e.g., "I believe most of my neighbours will adopt such a technology" (Lazaro et al., 2020; White et al., 2009). Social Identification was measured with a pictorial representation to assess the relationship with the community, based on the "Inclusion of the Other in the Self" scale (Aron et al., 1992; Gächter et al., 2015), ranging 5 points. Trust in People was measured using two items, e.g., "Generally speaking, most people can be trusted" (European Value Study (EVS); Caferra et al., 2021), and Trust in State with two items, e.g., "Please rate how much you trust in your legal system" (EVS; Caferra et al., 2021).

---

[4] Together with the device image, we randomly assigned participants to a control group, and groups with financial and environmental (CO2) savings information, and a group that was presented an app that would facilitate control of the smart device. We did not find any significant differences between these groups on any of our model or dependent variables and therefore will not discuss this intervention further.

[5] The cut-off criterion is number of items until the first item was reverse recoded.

[6] Based on an A-priori Sample Size Calculator for Structural Equation Models, with the specifications of our model, with a desired statistical power level of 0.8 and an estimated effect size of 0.2, we needed N = 425 observations to find an effect.

## 4.3. Data & statistical procedure

Data was handled with R statistics (R Development Core Team, 2008). Before conducting the analyses, all variables were mean-centered. Descriptive statistics are reported of raw scores. Building on a recommended two-step approach by Anderson & Gerbing (1988), we first conducted a Confirmatory Factor Analysis (CFA) before structural modelling to assess the fit of the measurement model. This step aims to estimate the measurement relationships between the observed variables and their underlying latent variables that cannot be directly assessed. The latent variables in our measurement model are behavioral intention (BI) as dependent variable, and performance expectancy, price value, effort expectancy, social norms, trust in state and trust in people, collective efficacy, and social identification.

To examine the psychometric properties of our measurement model, we used indicator reliability, internal consistency reliability, convergent validity, and discriminant validity as quality criteria. To test our hypotheses, a SEM was then calculated, which through the weighting of factors makes it possible to quantify the strength of the relationship between the latent variables.

---

[7] We calculated the model with and without trust in people, which did not change the results.

## 5. RESULTS

The measurement model consists of the latent variables and their underlying scale items for observation. Psychometric properties of our latent variables including item loadings can be found in the supplementary material (SM2). All loadings exceeded the recommended threshold of 0.7 on their respective scales except the first 'trust in people' item. This is further reflected in the results, as both McDonalds omega ($\omega$, McDonald, 1999) and Cronbach's alpha ($\alpha$, Cronbach, 1951) exceeded the recommended threshold of 0.7 for all cases, which confirms internal consistency reliability, except for trust in people. The same is true for the average variance extracted (AVE), where all values exceeded the recommended threshold of 0.5 (ranging from .69 to .90), while trust in people just met the minimum criteria with an AVE of .50. We will therefore interpret findings regarding trust in people with caution. Nevertheless, as we assessed trust in people through a well-established scale (based on the EVS), we decided to keep it in our model[7].

### 5.1. Psychometric properties of the measurement scales.

Building on recommendations of Hair et al. (2018), we assessed factor loadings for our constructs, as well as reliability indicators and mean variance extracted of items loading on the respective construct. We summarized these results for all model components in the supplementary material (SM2). Overall, the reliability of the scales was strong, with Cronbach's $\alpha$ ranging from .87 to .94 for key constructs such as Behavioral Intention, Price Value, Effort Expectancy, and Performance Expectancy. The Average Variance Extracted (AVE) values were generally high (.62-.80), indicating good convergent validity for most scales, though Trust in People had a lower reliability (Cronbach's $\alpha$ = .62, AVE = .49). To evaluate overall model fit, we used two absolute fit indices (RMSEA, SRMR) and two incremental fit indices (CFI, TLI)[8] which are all recommended for models with large sample sizes (Hair et al., 2018). The fit indices for both our CFA model and SEM model showed a good fit, as depicted in Table 1: All parameters of goodness exceeded the pre-defined cut-off, based on Hair et al. (2018).

---

[8] RMSEA – Root Mean Square Error of Approximation; SRMR – Standardized Root Mean Square Residual; CFI – Comparative Fit Index; TLI – Tucker Lewis Index

Table 1: CFA and SEM fit results, including fit cut-off.

| Measure | CFA | SEM | Cutoff |
|---|---|---|---|
| CFI | 0.928 | 0.922 | > 0.9 |
| TLI | 0.907 | 0.906 | > 0.9 |
| RMSEA | 0.077 | 0.071 | < 0.08 |
| SRMR | 0.036 | 0.043 | < 0.08 |

For discriminant validity, we found that in most cases, both the Fornell-Lacker Criterion (Fornell & Larcker, 1981) and in all cases, the conservative Hetereo-trait-mono-method (HTMT) criterion (Henseler et al., 2015) were met: The AVE values of each construct (in cursive) were higher than their squared correlations and the inter-construct correlations are below .85. Only for performance expectancy, we found an AVE which is below the squared correlations with behavioral intention, price value and collective efficacy. Still in this case, the conservative HTMT criterion was met (Henseler et al., 2015), which is why we accept discriminant validity to be given. Results are detailed in SM4.

## 5.2. Descriptive Results

Descriptive statistics of raw scores for the key constructs are reported in Table 2. BI to accept the smart heating technology was not normally distributed (Shapiro-Wilk-test: p < .001), with a mean value of 5.44 (SD = 1.51) in the overall sample[9]. On average, participants' ratings for both social and smart constructs of our model were quite high. We found the lowest average ratings for social norms. Generalized trust in people, and effort expectations were also above the scale mid-point. Social identification scores were not as high, with on average 2.4 on scale of 1 to 5.

Table 2: Descriptive statistics of raw scores.

| Scale | α | M | SD |
|---|---|---|---|
| Behavioral Intention | .94 | 5.44 | 1.51 |
| Price Value | .88 | 5.08 | 1.49 |
| Effort expectancy | .90 | 5.33 | 1.43 |
| Performance expectancy | .87 | 4.95 | 1.37 |
| Social norms | .87 | 4.35 | 1.50 |
| Trust in people | .62 | 5.18 | 1.98 |
| Trust in state | .82 | 5.01 | 2.33 |
| Social identification[1] |  | 2.39 | 1.09 |
| Collective efficacy | - | 5.03 | 1.49 |

---

[9] As data cannot be assumed to be drawn from a normally distributed population, we calculated all models (CFA & SEM) with a robust estimator, but find no differences in results.

1 Note that 'social identification' was assessed on a scale from 1-5

Country-specific means and standard deviations for they key constructs can be found in SM5. We find a significant difference between countries for BI [$F(4, 2245) = 46.9$, $p <.001$]. Post hoc comparisons using the Tukey HSD test indicated that the BI mean score for Estonia was significantly higher than for Austria ($p < .001$, 95% C.I. = [.21, .72]), Belgium ($p < .001$, 95% C.I. = [.35, .87]) and Germany ($p < .001$, 95% C.I. = [.25, .77]), and BI mean score for Greece was significantly higher than for all other countries (Greece – Austria: $p < .001$, 95% C.I. = [.73, 1.26]; Greece – Belgium: $p < .001$, 95% C.I. = [.87, 1.41]; Greece – Estonia: $p < .001$, 95% C.I. = [.27, .77]; Greece – Germany: $p < .001$, 95% C.I. = [.76, 1.31]). Further, Greek participants showed the highest average ratings particularly for individual beliefs, including price value, performance expectancy and effort expectancy. Higher age was a negative predictor for BI ($β = -.10$, $p = .008$). We found no differences between genders for smart heating technology acceptance ($β = -.04$, $p = .480$).

## 5.3. Structural model

Behavioral intention in our structural model had an $R^2$ value of .707, which exceeded the cutoff value of 0.10 for acceptable explanatory power for endogenous variable (Falk & Miller, 1992). $R^2$ values of all predictor items, reflecting the variance explained by the corresponding latent variable, were > 0.5, except for trust in people with .246 Fit indices for our model confirmed a good fit and Figure 2 shows the model results.

As depicted in Figure 4, for the overall sample, we found that price value had the strongest positive influence on BI, followed by performance expectancy and effort expectancy, in line with H1, H2 and H3. Of all our included social predictors, only trust in state had a small negative effect with higher trust levels reducing the intention to adopt, contradicting our hypotheses H4, H5 and H6. Additionally, also contrary to our hypothesis H4, we found a small negative interaction between social identification and social norms: social norms had a stronger effect on BI when people felt less close to their neighbors.

To examine differences between national samples, we calculated results grouped by country (A detailed overview of results is in SM7). Across all five groups, the general pattern was consistent, with strong significant effects of price value and effort expectancy. Performance expectancy was a significant predictor of BI for samples from Germany, Estonia and Belgium, but not for those from Austria and Greece. Additionally, collective efficacy had a significant positive effect on BI for Austrian participants; trust in state was a negative predictor for BI for participants from Greece. Finally, we found that the negative interaction between social norms and social identification was only found for our sample from Estonia, and the interaction was not significant for participants from the other countries.

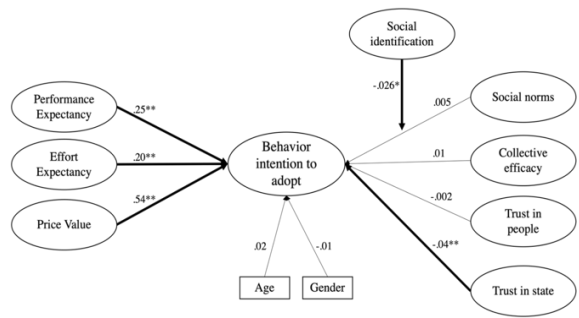

Figure 4: Model results for SEM.** $p < 0.001$; * $p < 0.05$

## 6. DISCUSSION

By means of structural equation modelling, we studied predictors of the intention to adopt a smart heating technology. The model included the technologies' perceived effectiveness, price value, effort expectancy as well as social beliefs including social norms, collective efficacy and different types of trust. Across five countries, the results indicate that the individual beliefs of the UTAUT model are suitable to predict the acceptance: price value, performance expectancy and effort expectancy were the most relevant predictors. Of the included social predictors, only trust in state had a small negative effect, and we found a small negative interaction between social identification and social norms. Within the specific country samples, some social predictors reached significance (e.g., collective efficacy in Austria) but overall, estimates were very small. For the aggregated model, we therefore can only accept H1 – H3, while hypotheses H4 – H6 must be rejected.

This indicates that individual beliefs currently better predict the intention to adopt smart heating thermostats: for the overall sample, and within the country samples, particularly financial aspects and technology related beliefs (usefulness, ease of use) influenced the intention to adopt a smart heating devices, in line with findings from prior research (Ahn et al., 2016; Girod et al., 2017; Tu et al., 2021).

Some of our results do not replicate evidence from previous similar studies, though. For example, social influence was the strongest predictor for the intention to adopt a smart thermostat in a representative consumer study in Germany (Große-Kreul, 2022), but we do not find any significant results for social beliefs – neither in the overall sample, nor in the representative German sample. One possible explanation lies in how social influence is operationalized: as we already discussed, the inconsistent results from social influence or social norms might be driven by whether the concept is understood as an injunctive or descriptive social norm, and who is considered within as norm-related group. Compared to Große-Kreul (2022), who assessed social influence as usage in other people and media presence (see Table 2), we used perceived descriptive social norms within the neighborhood.

We included collective efficacy, and the moderating effect of social identification, to broaden the interpretation of social influence as a singular construct, based on models and research of social influences on pro-environmental behavior. Although we did not find effects for most proposed social indicators in the overall sample, we found a small effect of collective efficacy in Austria. This might be a promising start for future research. In general, future work should consider the identified differences between countries, and gain a deeper understanding of this variations. Interestingly, we further found a negative interaction between social identification and social norms, which contradicts most earlier research (Cialdini & Jacobson, 2021). This might be driven by our choice of instrument: We used 'closeness of relations' within the neighborhood as indicator for social identification, which correlates highly with other relationship indicators, including knowledge of others' goals (Gächter et al., 2015). This better knowledge in turn might have limited the variability of perceived norms and therefore its potential to predict intentions. The effect size of the moderation effect is very small and country-level analysis shows it to be based in the Estonian sample.

### 6.1. Limitations

Firstly, we investigate how social beliefs relate to smart heating technology adoption, but causal conclusions can only be drawn within the limits of the SEM methodology we used; our assumptions about the causal impact of social belief predictors are here not supported by the empirical data, while our assumptions about the causal impact of individual predictors find support in the associations within the data, in line with prior empirical evidence.

Secondly, we did not measure adoption behavior, but rather the *intention* to adopt. The validity of the findings might be affected by the intention-behavior gap, which is found widely in pro-environmental consumer behavior (Carrington et al., 2010; Sheeran & Webb, 2016). This reflects in our data in the sense that we find very high adoption intention for smart technology, but the real adoption rate of smart thermostats in included countries has not yet reached full potential[10].

Lastly, the construct 'trust in people' was not found to exceed critical thresholds, e.g. the AVE and reliability criteria in our model, and findings regarding it should be interpreted with caution. Despite this, our CFA and SEM demonstrated a good model fit and almost all constructs met reliability and validity criteria. Thus, as we used a well-established scale from the EVS we decided for its inclusion in the final model.

## 6.2. Conclusion and Policy Implications

Overall, we conclude that the UTAUT model is well suited to explain behavioral intention to use smart thermostats. Our data did not yield support for an extension of the UTAUT model to include social beliefs derived based on evidence from previous pro-environmental behavior research; however, this does not indicate that they don't play a role. Possible explanations are the lack of publicness of thermostat adoption and current marketing practices focusing mainly on individual benefits. Social effects on pro-environmental choices are stronger on highly visible behaviors: studies find that visibility increases the perception of social status (Uren et al., 2021), collective efficacy affects public, but not private pro-environmental behaviors (Hamann & Reese, 2020), and people imitate visible behavior more (Babutsidze & Chai, 2018). Visibility has also been found to strengthen the relationship between a pro-environmental social identity and behavioral engagement (Brick et al., 2017). The adoption of a smart thermostat is invisible behavior, conducted in the private domain, so this might be a reason why social beliefs do not impact it greatly. The proposed model would benefit from implementation of explicit comparisons of private and public pro-environmental target variables to differentiate the effects of social beliefs, specifically with a focus on adoption of novel technologies.

The conceptualization of smart thermostat adoption as private sphere behavior also delivers a potential explanation of the negative significant effect of trust in state. As part of a meta-analysis, trust in state has been found to correlate with public pro-environmental behaviors (Cologna & Siegrist, 2020), and a study which examined the effect of generalized trust and trust in governments found that private behaviors are negatively correlated to trust in governmental institutions (Taniguchi & Marshall, 2018). This is explained with a theory of overreliance on the state, which decreases the perceived need or responsibility for own environmental action. It also seems worth investigating how the adoption and use of thermostats is framed across marketing campaigns and public service announcements. In a previous extensive qualitative analysis of product reviews for five commercial smart thermostats, technology or comfort related content categories were dominant (Malekpour Koupaei et al., 2020). The marketing emphasizes individual benefits, i.e. costs savings, energy efficiency and technology features; this might be one reason why social beliefs are not prevalent in people's cognitions regarding these devices. We second suggestions by for exampleLi's review on smart home adoption (Li et al., (2021)), that advertisements for such energy-efficiency devices should consider including broader social benefits, especially in light of potentially existing rebound effects (Dütschke et al., 2018; Seebauer, 2018).

Finally, studies on saving devices and efficiency technology often examine single individual's intention to adopt them. However, decisions about thermal comfort often rely on household decisions (Sintov et al., 2019). Sovacool et al. (2020) specifically identified decision-making structures around smart heating in households, displaying conflicts between different household members including between partners, roommates or parents and children. Care should be taken when interpreting results from our and previous literature about heating technology adoption based on individual's reported intentions, as in most cases (and in 94% of our sample), 'a household is not a person' (Seebauer & Wolf, 2017). Future studies should design measurements of target behaviors that are sensitive to both individual and household-level decision-making. Taken these findings into account, it is imperative to design interventions to better study how to accelerate the diffusion of smart heating technologies across Europe to the extent envisioned by policymakers.

---

[10] https://interpret.la/smart-home-sees-significant-growth-in-western-europe/

The data, analysis scripts and questionnaire with stimulus materials can be downloaded at https://osf.io/ba2vf/?view_only=986065e170584cad9098d0a2937e216b

# ACKNOWLEDGEMENTS

This research was supported by a European Union Horizon 2020 research and innovation programme grant (RENergetic, grant N957845; DECIDE, grant N894255) awarded to MB, FK, SK, CK. Conflicts of interests: none. MB: Conceptualization, Methodology, Data curation, Formal analysis, Visualization, Writing – original draft. FK: Conceptualization, Methodology, Investigation, Funding acquisition, Writing - review & editing. SK: Conceptualization, Methodology Writing - review & editing. CK: Conceptualization, Data curation, Formal analysis, Writing - review & editing, Supervision, Funding acquisition.